# Model and Performance of a No-Reference Quality Assessment Metric for Video Streaming

Mirghiasaldin Seyedebrahimi, Colin Bailey, Xiao-Hong Peng, *Senior Member, IEEE*

*Abstract*— Video streaming via TCP networks has become a popular and highly demanded service, but its quality assessment in both objective and subjective terms has not been properly addressed. In this paper, based on statistical analysis a full analytic model of a no-reference objective metric, namely Pause Intensity, for video quality assessment is presented. The model characterizes the video playout buffer behavior in connection with the network performance (throughput) and the video playout rate. This allows for instant quality measurement and control without requiring a reference video. Pause intensity specifically addresses the need for assessing the quality issue in terms of the continuity in the playout of TCP streaming videos, which cannot be properly measured by other objective metrics such as PSNR, SSIM and buffer underrun or pause frequency. The performance of the analytical model is rigidly verified by simulation results and subjective tests using a range of video clips. It is demonstrated that pause intensity is closely correlated with viewers' opinion scores regardless of the vastly different composition of individual elements, such as pause duration and pause frequency which jointly constitute this new quality metric. It is also shown that the correlation performance of pause intensity is consistent and content independent.

## I. Introduction

Video-on-demand services that utilize TCP, such as BBC's iPlayer, YouTube and Ustream, are rapidly growing with a prediction that Internet video traffic will account for 55 percent of all consumer Internet traffic in 2016 [1]. These services normally require a playout buffer to deal with the problems caused by TCP's congestion control mechanisms. Due to network bandwidth scarcity and the demand for high definition quality video buffer underrun can occur, which results in a pause of a certain length before video playback can resume when sufficient data has been received in the buffer. Although TCP is designed to guarantee the reception of all packets to ensure the image quality, buffer underrun will cause impairments in video playout continuity which can affect the viewer's perceived quality.

Developing proper quality metrics is currently one of the research focuses on video technologies and services. However, the need for assessing the impairment in playout continuity has not been fully addressed. This quality issue is important for video streaming service providers to monitor the viewer's quality of experience (QoE), especially in TCP based streaming as traditional metrics such as the peak signal-to-noise-ratio (PSNR) is unsuitable for quality measurement in this scenario [2]. There are some attempts to characterize buffer behavior with specific metrics, such as the buffer underrun frequency or probability [3]. However, these metrics are unable to demonstrate their correlation with subjective results in quality assessment, which is discussed further in Section II.

Our work builds upon a recently introduced new metric, namely *Pause Intensity* (PI) [4], and initial simulation and subjective testing results [5]. We show that viewer's QoE can be properly characterized by PI which is comprised of both pause frequency and pause duration, using the analytical model developed and extensive simulation and subjective testing results. We also establish the connection between PI and network performance such as throughput and service levels such as the video playout rate. This unique feature makes PI a reference free metric which can be used to enable adaptive traffic control in streaming service delivery to meet the quality requirement defined by a certain PI value.

Discussions on the related work are given in Section II. Sections III describes the characteristics of buffer underrun in the context of a streaming network. The analytical model of PI is derived and appraised in Section IV. In Section V the simulation and testing environments are set up, followed by the validation of the PI model by both simulation and subjective testing results in Section VI. The paper is concluded in Section VII.

## II. Related Work

Metrics used for measuring video quality have typically been classified into three categories: full reference (FR), reduced reference (RR) and no reference (NR). FR metrics employ the original video as a reference point and the quality is determined by computing the difference between the original and distorted videos. The most common FR metric used is PSNR which is directly derived from the mean squared error (MSE). Due to the limitation of PSNR [6] [7] many other metrics have been proposed, such as structural similarity (SSIM) [8], video quality metric (VQM) [9], and other licensed tools such as SwissQual [10] and Kwill [11].

NR metrics assess the content quality level without any

The authors are with the School of Engineering & Applied Science, Aston University, Birmingham B4 7ET, UK (email: seyedebm@aston.ac.uk; baileyc1@aston.ac.uk; x-h.peng@aston.ac.uk).

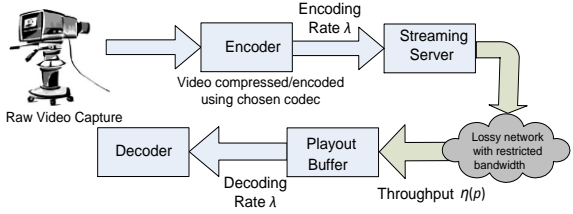

Fig. 1. Video streaming architecture

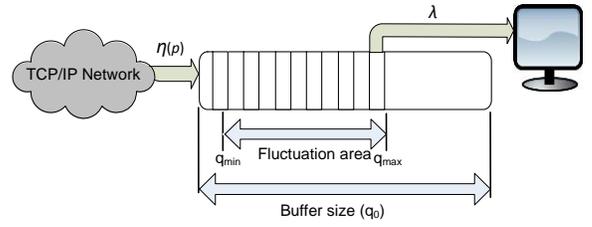

Fig. 2. Buffer structure with related settings.

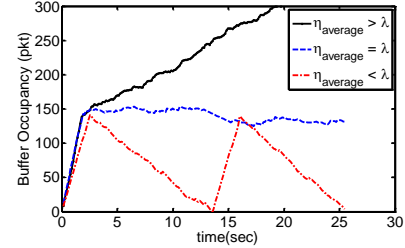

Fig. 3. Buffer characteristics

knowledge of the original material. Commonly, this is done by identifying artifacts such as blurriness, blockiness, sharpness or a combination of related artifacts. Ferzli [12] and Liu [13] show that their metrics on blurriness and blockiness, respectively, is highly correlated with subjective data. The above metrics characterize the typical examples of artifacts that occur in best effort networks where data delivery is not guaranteed. In a TCP based streaming session, however, the artifacts such as blockiness, blurring and sharpness are not typically experienced as data delivery is ensured.

The work we present falls into the NR category and is concerned with a video evaluation tool for TCP based streaming. Other studies in this category have investigated varied aspects in the buffer underrun phenomenon. Kim [3] proposed a model to find a desirable buffer size determined by network characteristics and the underrun probability. However, no satisfaction assessment is carried out apart from an assumption that a higher underrun probability can result in greater dissatisfaction.

The buffer starvation probability is a similar term to underrun probability, pause frequency, underflow probability or jitter frequency. It has been shown [14] that the starvation probability can be reduced by optimizing buffer settings and playout rate smoothing factors. The analysis of buffer delay has also been carried out, giving the conditions to avoid overflow and underflow in time varying channels [15] and the delay-underflow trade-off in a lossy network [16].

There are also schemes intended to control and reduce pause occurrence by using TCP-friendly stream rate adaptation to the changes in buffer occupancy [17] or optimizing TCP bandwidth allocation [18]. In [19] the buffer occupancy and network status have been used to evaluate and optimize the end user's perceived quality for video streaming, concerning the received video resolution levels but without considering continuity. The importance of mitigating buffer underrun in TCP based streaming has been highlighted in a recent survey paper [20].

Those metrics used to characterize buffer behavior including buffer-underrun or pause frequency and its related parameters fall short since they do not reflect viewer's QoE. This fact will be verified later in Section VI using subjective assessment results. We will also prove that the new objective assessment metric, pause intensity, can precisely quantify the buffer underrun effect on the viewer's perception of playout quality, and that its correlation with the viewer's QoE is content independent.

III. THE CHARACTERISTICS OF BUFFER UNDERRUN

In this section, the nature of buffer underrun and the resulting impairments are discussed. Buffer underrun is typically a result of inefficiency in the network due to bandwidth limitation and/or packet loss (e.g. TCP/IP over WiFi, congested network) which results in a throughput less than the required decoding rate of the video.

A pause is defined in this work as a temporary suspension of play followed by a period of playback which resumes from where the pause occurred. An overview of a typical streaming network is shown in Fig. 1. The video player (decoder) provides the sink for the outgoing packets. The rate of the successfully received data from the TCP connection or termed throughput, $\eta$, is a function of the packet loss probability, $p$, of the network. The required video playout rate, $\lambda$, is a characteristic of the video codec which is determined by the visual quality required of the video.

Our discussion is focused on the frequent buffer underrun phenomenon with different play-pause durations due to the throughput being less than the playout rate required of the video. Fig. 2 shows the structure of the playout buffer in connection with the network. The size of the playout buffer in the receiver is typically large enough to absorb the effect of small fluctuations in the network throughput and usually of sufficient size to store a few seconds worth of video data. The predefined threshold for video playback is referred to as $q_{max}$. Video playback will cease whenever the amount of stored packets in the buffer is less than the minimum threshold $q_{min}$. The packets are sorted using a FIFO algorithm. The instant occupancy of the buffer is referred to as $q$.

When the video is played, an initial delay is experienced from the moment when the buffer receives data until the buffer occupancy reaches $q_{max}$. Following the initial delay, playback starts and one of three scenarios may take place as illustrated in Fig. 3. Actually in the case where the average throughput is greater than or equal to the required playout rate $\lambda$, ($\eta_{average} \geq \lambda$), and provided that the receiver has enough memory to buffer the received data, there will be just an initial delay occurrence followed by a long period of video playback.

When the available network throughput is less than $\lambda$. Pause events will be experienced whenever the amount of data in the buffer falls below $q_{min}$. This is the basis for our analytical model presented in Section IV.

## IV. MODEL OF PAUSE INTENSITY

In this section, we show that the behavior of pause and play events during the playback follows the performance of throughput in a lossy channel. The characteristics of these events and throughput can be described using a unimodal probability density function (pdf) with their statistical moments (e.g. the mean values). Using the mean values we are able to develop a simple model for the playback buffer and derive the closed form formulations for pause duration, pause frequency and ultimately the pause intensity, as a function of throughput, video playout rate and receiver buffer settings.

### A. Throughput Characteristics

Pause occurrence during the playback of a video stream is mainly caused by insufficient incoming data rate of the buffer or network throughput due to packet loss. Packet loss as a stochastic process has been studied extensively in the context of TCP/IP network [21]-[23]. Since the detailed characteristics of the distribution of packet loss and throughput (e.g. average, skewness and kurtosis) depend on the network structure and traffic status, the following assumptions and considerations have been made for the subsequent discussions:

a) Bernoulli, Geometric and Gamma functions have been employed to fit the loss event distribution functions [21]-[23]. A continuous Gamma density function as the distribution function of 'probability of packet loss', $p_l$, is used in this work, so that any of the above mentioned functions could be utilized and compared. Fig. 4(a) shows an example of the assumed pdf of $p_l$, $f_P(p_l)$, with an average packet loss rate of 0.02 and generated using *Gamma* ($k=2.8$, $\theta=0.7$) where $k$ and $\theta$ are the shape and scale of a gamma distribution, respectively.

b) Although throughput monotonically decreases with respect to the packet loss rate in a TCP connection, the shape and pace of the change depend on the exact variant of TCP and the type of the network involved. For the rest of this work we adopt the widely accepted TCP-Reno model [24] and related specifications [25] for network throughput analysis. The throughput for a TCP-Reno connection is given as [24]:

$$\eta(p_l) = \begin{cases} \frac{1}{R\sqrt{\frac{2bp_l}{3}}+T_0 min\left(1,3\sqrt{\frac{3bp_l}{8}}\right)p_l(1+32p_l^2)} & \text{with timeout} \\ \frac{1}{R}\sqrt{\frac{3}{2bp_l}} & \text{without timeout} \end{cases} \quad (1)$$

where $T_0$ is the timeout in TCP-Reno and $R$ is the round trip time. The throughput model without the timeout effect in (1) is invertible but not sufficient to analyze the pause duration behavior. Therefore, the non-invertible model of throughput with the timeout effect will be used in the rest of the paper. The number of rounds for each increment in window size, $b$, in compliance with our simulation setting is assumed to be 2.

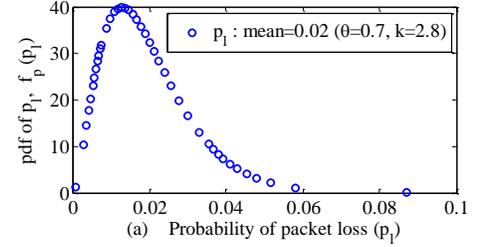

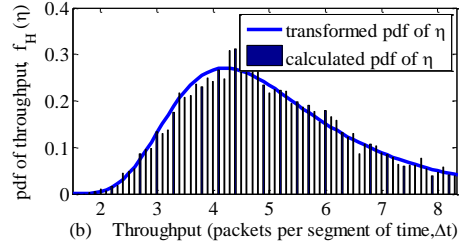

Fig. 4. Examined pdf of (a) the probability of packet loss and (b) achieved throughput.

Furthermore, the adequate range of $p_l$ lower than 0.12 will satisfy the condition to assume $min\left(1,3\sqrt{\frac{3bp_l}{8}}\right) = 3\sqrt{\frac{3bp_l}{8}}$. Hence the throughput concerned as a function of $p_l$ is:

$$\eta_{Reno}(p_l) = \frac{1}{\frac{2R}{\sqrt{3}}p_l^{\frac{1}{2}}+\frac{3\sqrt{3}T_0}{2}p_l^{\frac{3}{2}}(1+32p_l^2)} \quad (2)$$

The actual throughput of a TCP connection is also affected by the advertised window setting from the receiver side and bandwidth limitation due to a bottleneck in this connection. This will be considered later in the model formulation and evaluation.

c) To find out the transformed pdf of a random function (i.e. throughput, $\eta$) based on the known pdf of its random variable (i.e. probability of packet loss, $p_l$), the following relation will be used when the function is invertible.

Given a function $y=g(x)$, if the distribution function of $x$ is known as $f_x(x)$, then the distribution function of $y$ is given by

$$f_Y(y) = \frac{f_X(g^{-1}(y))}{|g'(g^{-1}(y))|} \quad (3)$$

A variation of (3) in the case of $y$ being a non-invertible function is

$$f_Y(y) = \sum_i \frac{f_X(x_i)}{|g'(x_i)|} \quad (4)$$

in which $x_i$ is the $i$-th root of $y=g(x)$ for a given value of $y$. This relation can be applied to finding the pdf of throughput, $\eta(p_l)$, based on the known pdf of $p_l$ without an explicitly inverted $p_l(\eta)$. Since $\eta$ monotonically decreases with respect to $p_l$, it is possible to solve (2) numerically for a given value of $\eta$ and find the corresponding pdf, $f_H$, using (4). Fig. 4(b) shows the achieved pdf of throughput for the given pdf of the packet loss probability. To evaluate the accuracy of this approach, a histogram of 5000 calculated values of throughput is shown alongside. Throughput values are calculated based on random packet loss rates derived from the given Gamma pdf. The comparison shows a good compliance and an increase in accuracy when the number of samples is large.

## B. Playback Characteristics

The achieved distribution of throughput will be used to analyze the pause behavior as follows. In Fig. 5 the instant growth of buffer occupancy, $\Delta q_i$, and the total occupancy, $q$, can be written as a function of throughput during the time interval $\Delta t_i$, i.e.:

$$\Delta q_i = \Delta t_i . \eta_i \quad \text{and} \quad q = \sum \Delta q_i = \sum \Delta t_i . \eta_i \quad (5)$$

where $\eta_i = \eta(p_i)$. If $\Delta t_i$ is constant i.e. $\Delta t_i = \Delta t$, and small enough to assume more than one segment per each pause event, then the occupancy of buffer, $q$, after $m$ segments will be a random variable equal to the summation of $m$ random variable, i.e.:

$$q = \Delta t \sum_{i=1}^{m} \eta_i \quad (6)$$

A value of $\Delta t = 100ms$ will be sufficient in (6) to satisfy the above conditions. If $q_0$ is the difference between the minimum threshold (i.e. the trigger of pause) and the maximum threshold (i.e. the trigger of play resumption) of the buffer, finding the value $m = m_0$ that leads to $q = q_0$ is actually another way of presenting the pause duration, $v$, i.e.:

$$\begin{cases} v = m_0 \Delta t \\ where\ m_0\ satisfies:\ \Delta t \sum_{i=1}^{m_0} \eta(p_i) = q_0 \end{cases} \quad (7)$$

The equation (6) is a summation of $m$ values of a random variable with the known derived pdf described in the previous subsection. The $m$ variables in (6) have an independent and identical distribution (i.i.d). As long as the assessment segment $\Delta t$ is small enough, compared to the minimum possible length of a pause to guarantee a large value of $m$ (e.g. $m>10$), the central limit theorem can be used to find the distribution of $q$ in (6), i.e. for $\eta: f_H(\mu_\eta, \sigma_\eta)$, hence we have

$$q: \mathcal{N}_Q(m\mu_\eta, \sqrt{m}\sigma_\eta) \quad (8)$$

where $f_H$ is the pdf of throughput with mean and variance $\mu_\eta$ and $\sigma_\eta$, respectively (referring to Fig. 4). The pdf of buffer occupancy, $N_Q$, is a normal distribution with a mean equal to $m\mu_\eta$ and a variance equal to $m\sigma_\eta^2$. Obviously a subset of $N_Q$ values, which are relevant to $q = q_0$ when $m = m_0$, will satisfy (7). Given the relation between $m_0$ and $v$ in (7), these values will lead to the desired values for pause duration $v$, and their probabilities. Therefore, the probability of occurrences of the pause duration $v$ can be expressed as:

$$\begin{cases} p(v) = p(q = q_0)|_{m=m_0} = \frac{1}{\sigma_\eta \sqrt{2\pi m_0}} e^{-\frac{(q_0 - m_0 \mu_\eta)^2}{2m_0 \sigma_\eta^2}} & \forall\ m_0 > 0 \\ given\ \eta: f_H(\mu_\eta, \sigma_\eta),\ q: \mathcal{N}_Q(m\mu_\eta, \sqrt{m}\sigma_\eta) \end{cases} \quad (9)$$

From (7), (8) and (9), we can conclude that for any given pdf of $\eta$ with its $\mu_\eta$ and $\sigma_\eta$ the probable pause durations can be described as a subset of normal distribution given in (8), regardless of the type of distributions for $\eta$. It is noted that (9) includes some samples of (8) which satisfy the condition given in (7). Fig. 6 depicts (9) as the examples of $p_l$ and $\eta$ in Fig. 4.

During the pause time there is no packet consumption in the buffer, so $\lambda$ has no influence during this period. Using a similar method we can derive the distribution of play duration, $v'$, shown in Fig. 5, during which both throughput and the

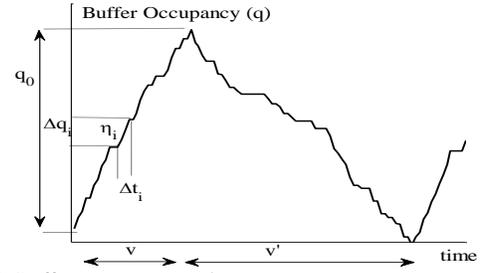

Fig. 5. Buffer occupancy vs. time.

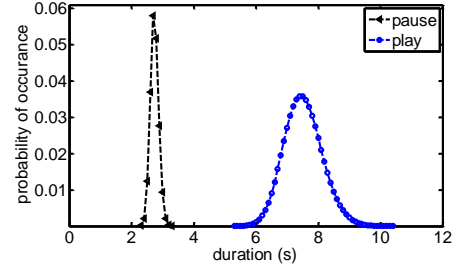

Fig. 6. Distributions of pause and play durations

playout/code rate ($\lambda$) must be considered while $\lambda$ is assumed to be constant, i.e.:

$$\begin{cases} v' = m_1 \Delta t \\ where\ m_1\ satisfies\ \ \Delta t \sum_{i=1}^{m_1}[\eta(p_i) - \lambda] = q_0 \end{cases} \quad (10)$$

Fig. 6 shows the probability of occurrence of the play duration corresponding to the pause duration derived previously.

Although the above discussion gives an insight into the behavior of pause due to the impairment of transport connections, it does not give a direct closed form expression showing the relationship between the pause duration or pause frequency and network conditions such as throughput or the packet loss rate. Most importantly, it does not show how this pause or buffer behavior affect the quality perceived by the end users of video streaming services using the TCP protocol. In the next subsection, the analytical model of the quality metric, Pause Intensity, will be presented, which combines the statistical features of both pause duration and pause frequency.

## C. Pause Intensity Model

By applying the transformed pdf and the central limit theorem we have shown that throughput, pause duration and play duration have unimodal distributions. This allows us to use the statistical elements such as the mean or maximum probable value to achieve the closed form representations of buffer characteristics. Any other parameter such as pause frequency will be defined as a function of these representative values. In this subsection, we establish the model of pause intensity, a no-reference quality assessment metric, based on the buffer underrun properties for video streaming in TCP networks. These buffer underrun properties are characterized by the average pause duration $\bar{v}$, and average pause frequency, $\bar{f}_v$, given network and buffer conditions and settings such as throughput, the code rate and receiver buffer fluctuation area.

Fig. 7 shows a typical pause-play period with all related parameters, where the playout buffer is large enough to absorb the effect of small fluctuations. The trend of the accumulated data in the buffer between $t_{v1}$ and $t_{max}$ follows a line with the gradient equal to the average throughput in that period. The durations of pauses and plays are denoted by $v$ and $v'$, respectively, and $w$ represents the duration of a pause-play event. A pause event occurs when the number of buffered packets is reduced to $q_{min}$ at time $t_{v1}$. Play will resume whenever the number of buffered packets reaches $q_{max}$ at time $t_{max}$. If the average throughput is less than the required video playout rate $\lambda$, the next pause will occur at time $t_{v2}$. The buffered data during the pause-play event can be expressed as:

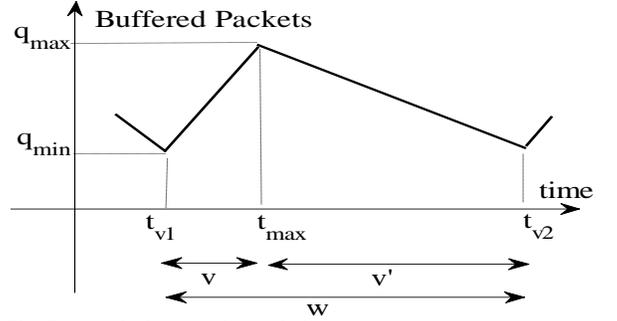

Fig. 7. A typical pause-play period

$$\begin{cases} q_{pause} - q_{min} = \frac{q_{max}-q_{min}}{t_{max}-t_{v1}}(t-t_{v1}), & t_{v1} < t \leq t_{max} \\ q_{play} - q_{max} = \frac{q_{max}-q_{min}}{t_{max}-t_{v2}}(t-t_{max}), & t_{max} < t \leq t_{v2} \end{cases} \quad (11)$$

Unlike the play time, there is no output from the buffer during the pause, which leads to:

$$\begin{cases} q_{pause} = q_{min} + \eta(t - t_v), & t_{v1} < t \leq t_{max} \\ q_{play} = q_{max} + (\eta - \lambda)\cdot(t - t_{max}), & t_{max} < t \leq t_{v2} \end{cases} \quad (12)$$

From (11) and (12), the parameters $v$, $v'$ and $w$ can be expressed as:

$$\begin{cases} v = t_{max} - t_{v1} = \frac{q_0}{\eta} \\ v' = t_{v2} - t_{max} = \frac{q_0}{\lambda-\eta} \\ w = \frac{q_0\lambda}{\eta(\lambda-\eta)} \\ q_0 = q_{max} - q_{min} = flactuation\ area\ in\ the\ buffer \end{cases} \quad (13)$$

Recalling the definitions of pause intensity, the average pause duration and pause frequency from [4] and discussions in Subsection IV-C, we use the average values to represent the parameters given in (13), and then we have.

$$\begin{cases} average\ pause\ duration = \bar{v} = \frac{q_0}{\eta} \\ pause\ frequency = \bar{f}_v = \frac{1}{\bar{w}} = \frac{\eta(\lambda-\eta)}{q_0\lambda} \\ Pause\ Intensity = PI = \bar{v}\cdot\bar{f}_v = 1 - \frac{\eta}{\lambda} \end{cases} \quad (14)$$

The pause duration $\bar{v}$ is a function of a dedicated buffer fluctuation area $q_0$ and network throughput $\eta(p)$. It is however not affected by the video playout rate $\lambda$ as no output from the buffer during the pause. Pause intensity (PI) represents the relative effectiveness of throughput compared to the required playout rate $\lambda$ and is not affected by $q_0$. The frequency of a pause-play sequence $\bar{f}_v$ is built upon the combination of the playout rate, network throughput and buffer settings (i.e. the buffer size). These features will be exploited further during the performance analysis in the following sections.

As explained in Section IV-A, we adopt TCP-Reno in our work and the throughput used in the PI model (14) is given in (2). Recalling the nature of a pause-play period $w$, intuitively it can be seen from Fig. 7 that $w$ has a finite value if $\eta_{average} < \lambda$. The change rate of $w$ in (13) is given by:

$$dw = \frac{\partial w}{\partial \eta}d\eta + \frac{\partial w}{\partial \lambda}d\lambda = q_0\left(\frac{-\lambda(\lambda-2\eta)}{\eta^2(\lambda-\eta)^2}d\eta + \frac{-1}{(\lambda-\eta)^2}d\lambda\right) \quad (15)$$

Furthermore, the change rate of $w$ with respect to $\eta$, defined as $\beta$, can be determined by:

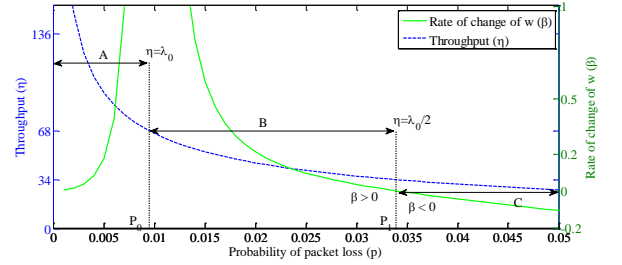

Fig. 8. Critical points of pause-play sequence in relation to throughput and packet loss probability.

$$\begin{cases} \beta = \frac{dw}{d\eta} = q_0\frac{-\lambda(\lambda-2\eta)}{(\eta(\lambda-\eta))^2} \\ \lambda = \lambda_0 = cte,\ \eta = \min\left(\eta_{Reno}(p), BW_{min}, \frac{W_m}{RTT}\right) \end{cases} \quad (16)$$

in which the playout rate $\lambda$ is considered to be a positive constant value without a continuous change. The actual throughput $\eta$ is the minimum among the calculated throughput in (16), the bottleneck bandwidth $BW_{min}$ and the advertised window size per round-trip time $W_m/RTT$.

Fig.8 depicts the features given by (16) and the critical points against the probability of packet loss rate $p$. As it is shown, the vertical asymptote $p= P_0$ in which $\eta(p_0) = \lambda_0$ is the returning point of $w$ from infinity, and consequently it is the point from which pauses begin to occur within the range denoted by $B$. From the point where $p=P_1$, which results in $\eta(p_1) = \lambda_0/2$, the pause duration starts exceeding the play duration and viewers satisfaction will be less likely within the range denoted by $C$. Later we will see that $p=P_1$ is the extremum point of pause frequency as well. Throughput in the range denoted by $A$ is higher than what is required and pauses are therefore unlikely to happen.

The pause intensity model established in this section will be validated by simulation and subjective testing to show its effectiveness, accuracy and other features in the following sections.

## V. SIMULATION AND SUBJECTIVE TESTING

### A. Simulation Setup

The simulation was carried out using NS-2 with the parameters specified in Table I. A simple bottleneck was established with a single sender and receiver. Only one video stream was assumed and no background traffic was employed. Packet loss was set to vary from 0% to 12% which affects network throughput and consequently buffer behavior. An element of randomness was added into the timing of packet loss occurrence, i.e. for each loss rate, 10 simulations were executed such that the packet loss timing was varied with the level of buffer occupancy. The mean and deviation of the simulation results for that loss rate can then be shown.

TABLE I
SIMULATION SETUP

| Connection | TCP(Reno) | Video code rate $\lambda_0$ | 100 KB/s |
|---|---|---|---|
| Bottleneck | 1Mb/s | Packet Loss Range | 0%-12% |
| Packet size | 1500 B | $q_{max}$ | 200KB |
| $RTT_{average}$ | 128 ms | $q_{min}$ | 1.5KB |
| $T_0$ | 128 ms | Window Size $W_m$ | 20 |

TABLE II
SUBJECTIVE TESTING SETUP

| | Subjective Testing 1 | | | | Subjective Testing 2 |
|---|---|---|---|---|---|
| | MotoGP (M) | Run (R1) | News (N) | Cartoon (C) | Rally (R2) |
| codec | H264 | H264 | H264 | H264 | H264 |
| resolution | 540x360 | 640x360 | 640x360 | 640x360 | 640x360 |
| frame rate | 30 | 25 | 25 | 25 | 30 |
| encoding rate | 840kbps | 781kbps | 781kbps | 781kbps | 788kbps |
| video length | 90s | 90s | 90s | 90s | 90s |
| reviewers/ clip | 19 | 17 | 16 | 17 | 20 |
| no. of clips | 16 | 10 | 10 | 10 | 12 |

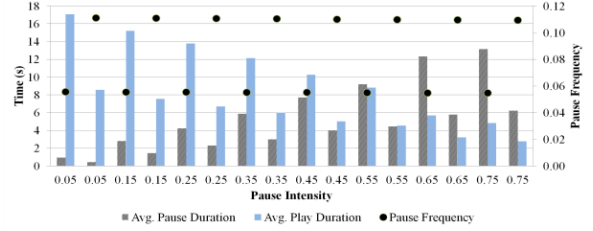

Fig. 9. An example of the characteristics for subjective testing-1 (MotoGP video sequence scenarios).

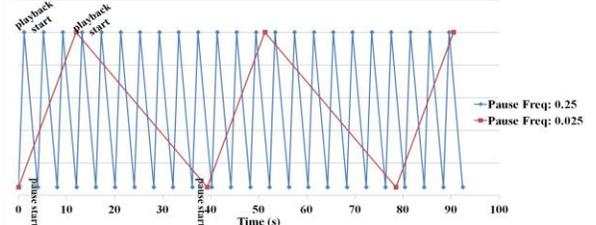

Fig. 10. Pause characteristics of two videos with very different frequencies but the same PI.

NS-2 was used to provide the output of the network and a separate module was written to simulate buffer behavior based on the discussion in Section IV. The buffer size (198.5*KB* or ~133 packets) was selected to be approximately double the video coding rate, which correlates to around 2 seconds of video. The playout rate and buffer size were also selected based on the requirements of live streaming services.

*B. Subjective Testing Setup*

In order to verify the success of the pause intensity metric, subjective testing was also carried out. Testers were instructed that ratings given for each video should represent their overall viewing experience and reflect their real viewing expectations. Due to the nature of the impairment, relatively long video sequences of 90 seconds were used. In accordance with ITU P.911 [26], a five scale overall quality rating, i.e. Mean Opinion Score (MOS) was employed with the terminology recommended by the ITU. In contrast to the previous work [5] which used the Degradation Category Rating (DCR), this work adopted the single stimulus method: Absolute Category Rating (ACR). Table II shows the corresponding setting parameters for two groups of subjective testing that were carried out: Subjective Testing-1 and Subjective Testing-2.

In Subjective Testing-1, four different types of video sequences were used, namely MotoGP, Run, News and Cartoon, with similar parameters as shown in Table II. 'MotoGP' have a large amount of fast paced motion in scenes and a large number of cuts from scene to scene. 'Run' contains a lot of motion in the image itself and, in addition to which there is a substantial amount of camera panning. 'News' on the other hand contains neither camera panning nor large amounts of motion in the clip and with few scene changes. 'Cartoon' contains a small amount of panning, a fair amount of movement within frames and a large number of scene changes. The choice of these clips was aimed to assess not only the correlation property of pause intensity (the new objective metric) with the subjective opinion scores, but also the content independency of PI.

In Subjective Testing-2, further investigation was carried out to assess the robustness of PI in terms of accommodating different buffer characteristics, especially to test if the metric still provides a good correlation with viewer experience in more extreme cases. This part of the work therefore provides stress testing evaluating the presence of either high pause frequency or long pause duration whilst maintaining a constant value of PI. The content named 'Rally' is chosen for the discontinuity to make an obvious contrast for testers. Characteristics of this video sequence are in the same range of Subjective Testing-1 given in Table II.

The detailed compositions of each PI used for both groups of subjective testing are provided in Tables III and V, respectively, in Section VI.

Fig. 9 shows an example of the impairment characteristics in Subjective Testing-1 using the MotoGP video sequences. In this case, varying levels of pause intensity were used in a range of 0.05 − 0.75. As discussed previously, each pause intensity value is repeated because the pause duration and frequency is varied to make different compositions of the same pause intensity value. To reveal the buffer behavior for each pause intensity value, two different compositions for the same PI value were used. The PI values shown on the horizontal axis of Fig.9 represent the first and second scenarios alternatively. The first scenario has a lower pause frequency (represented by the black points in Fig.9) than the second scenario, meaning that both the play duration and pause duration in first scenario are larger than those in scenario 2.

Fig. 10 shows an example of videos used in Subjective Testing-2 (stress testing of PI with either extremely high pause frequency or long pause duration), which illustrates the contrast between pause-play scenarios with pause frequencies being 0.03 and 0.25, respectively, but with both having the same pause intensity. The upward slope is the time taken to fill up the buffer and the downward slope shows the time when the video is played out. In addition, the significant difference in pause duration between the two videos is also clearly visible.

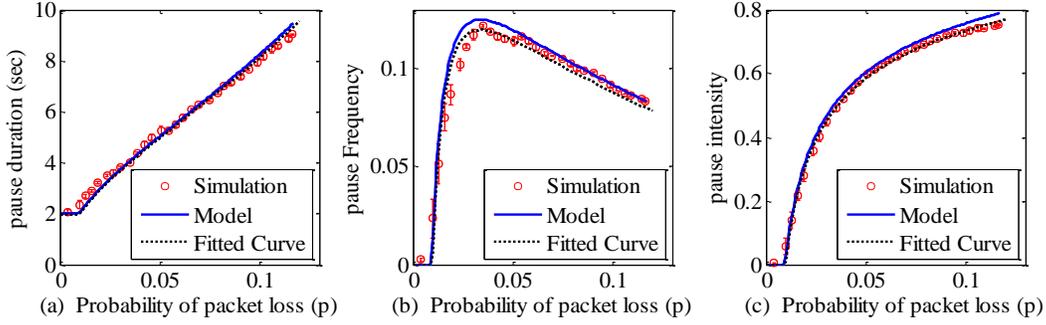

Fig. 11. Model and simulation comparisons: (a) pause duration vs. loss probability; (b) pause frequency vs. loss probability; and (c) PI vs. loss probability.

In all tests [27], video sequences with specified pause intensity were selected randomly from the pool of possible scenarios. Testers were allowed to vote for up to 6 clips in each run and each video was run between 14-20 times for the purpose of evaluation. Testers were invited through social media and local communities and largely did not possess any specific technical background. The testing environment also allowed the user to choose whether to continue to watch or not due to the level of impairment imposed.

## VI. RESULTS AND ANALYSIS

### A. Model Verification by Simulation

Fig. 11 illustrates the results of the analytical model in comparison with the simulation results obtained in the environment with the parameters given in Table I. As described earlier, each value produced in simulations represents the mean and deviation of 10 runs. Additionally, for each set of simulation test results, the best fitted curve to the model has been depicted. Clearly, both simulation and model results, as exhibited in Fig. 11, are closely matched, which suggests that our model can successfully characterize the buffer underrun behavior with a high precision.

Fig. 11(a) shows the changes in pause duration against the probability of packet loss in the network. It may be recalled that pause duration is a function of throughput $\eta$ and the buffer fluctuation area $q_0$, which is not dependent of the playout rate $\lambda$, as demonstrated in (14). It can be understood that as the packet loss probability increases, throughput decreases and therefore more time is required to fill the buffer to the playout threshold level $q_{max}$. It is also noticed that pause duration remains unchanged for the small values of $p$ as network throughput is relatively stable under this condition.

Another important point we would like to make is based on the result of pause frequency or underrun probability, as shown in Fig. 11(b). It reveals that pause frequency does not change monotonically as the packet loss probability increases. It increases with the loss probability when $p$ is relatively small, and will adversely decrease when $p$ increases up to a certain value (denoted by $P_1$ in Fig. 11(b) and Fig. 8). This is because the pause duration increases steadily with the loss probability as shown in Fig. 11(a). As a result, the number of pauses will be reduced, so will the pause frequency. This phenomenon reveals that using pause frequency or the underrun probability alone simply cannot reflect the video playout quality as the viewer's QoE is directly affected by the change in loss conditions of the network.

Fig. 11(c) shows the performance of pause intensity which changes monotonically in the whole range of the packet loss probability specified. The advantage of using PI is demonstrated as opposed to using a single parameter such as pause frequency as discussed above. PI shows a positive correlation with the packet loss probability, meaning that the metric is able to reflect network conditions, which is not the case for pause frequency and other parameters associated with buffer underrun such as average pause duration and average play duration. In summary, PI is a comprehensive metric that encompasses all the behavioral characteristics of the buffer. This conclusion will be further confirmed by subjective testing results discussed later.

### B. Subjective Assessment-1 for Pause Intensity

As described in Section V, 45 video sequences of four different types (MotoGP, Run, News and Cartoon) were used in the first group of subjective testing. The detailed setting parameters such as pause intensity, pause frequency and pause duration, and the testing results in terms of the MOS rating for each test are given in Table III. Fig.12, constructed using the information provided in Table III, demonstrates the advantages of using the pause intensity metric over pause frequency and pause duration.

First of all, Fig. 12(a) shows two important features of this new metric: (1) pause intensity is closely correlated with the viewer's experienced quality for a wide range of PI values which encompass varying compositions of pause frequency and pause duration; and (2) this correlation is consistent with different types of video sequences tested, or in other words, pause intensity is highly content independent.

In Figs. 12(b) and 12(c), the results show that both pause frequency and pause duration have poor correlation with MOS for all the video content used. For example from Table III, the pause frequencies of Videos 1 and 15 are identical, but their corresponding mean opinion scores (MOS) are 4.35 and 1.43, respectively. A similar result can be found by evaluating the average pause duration in Videos 6 and 15. In this case, the pause duration shows a variation of just 0.34 seconds, but the MOS varies greatly (3.42 and 1.43, respectively). Clearly, it is confirmed that pause frequency or pause duration alone does not provide a good correlation with viewer opinion.

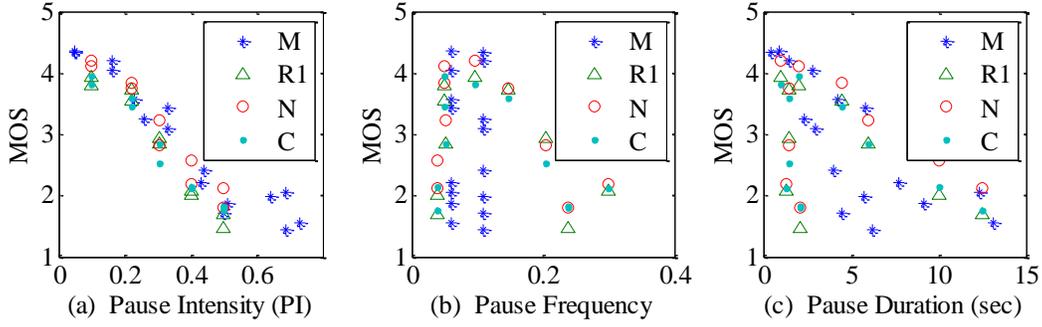

Fig.12. Results for Subjective Testing-1: MOS vs. (a) PI, (b) pause frequency, and (c) pause duration.

TABLE III
SUBJECTIVE TESTING-1 RESULTS

| Video ID | Video content | PI | Pause Frequency | Average Pause Duration (sec) | MOS |
|---|---|---|---|---|---|
| 0 | M | 0.05 | 0.06 | 0.93 | 4.35 |
| 1 | M | 0.05 | 0.11 | 0.43 | 4.34 |
| 2 | M | 0.16 | 0.06 | 2.8 | 4.03 |
| 3 | M | 0.16 | 0.11 | 1.46 | 4.19 |
| 4 | M | 0.23 | 0.06 | 4.22 | 3.57 |
| 5 | M | 0.26 | 0.11 | 2.31 | 3.24 |
| 6 | M | 0.33 | 0.06 | 5.87 | 3.42 |
| 7 | M | 0.33 | 0.11 | 3 | 3.08 |
| 8 | M | 0.43 | 0.06 | 7.72 | 2.21 |
| 9 | M | 0.44 | 0.11 | 3.97 | 2.41 |
| 10 | M | 0.51 | 0.06 | 9.2 | 1.88 |
| 11 | M | 0.5 | 0.11 | 4.47 | 1.72 |
| 12 | M | 0.69 | 0.06 | 12.33 | 2.05 |
| 13 | M | 0.64 | 0.11 | 5.79 | 1.99 |
| 14 | M | 0.73 | 0.06 | 13.16 | 1.55 |
| 15 | M | 0.69 | 0.11 | 6.21 | 1.43 |
| 16 | R1 | 0.097561 | 0.04878 | 2 | 3.8 |
| 17 | R1 | 0.097087 | 0.097087 | 1 | 3.933333 |
| 18 | R1 | 0.219512 | 0.04878 | 4.5 | 3.55 |
| 19 | R1 | 0.220588 | 0.147059 | 1.5 | 3.733333 |
| 20 | R1 | 0.305344 | 0.050891 | 6 | 2.85 |
| 21 | R1 | 0.306122 | 0.204082 | 1.5 | 2.933333 |
| 22 | R1 | 0.401606 | 0.040161 | 10 | 2 |
| 23 | R1 | 0.401198 | 0.299401 | 1.34 | 2.066667 |
| 24 | R1 | 0.5 | 0.04 | 12.5 | 1.7 |
| 25 | R1 | 0.5 | 0.238095 | 2.1 | 1.466667 |
| 26 | N | 0.097561 | 0.04878 | 2 | 4.11 |
| 27 | N | 0.097087 | 0.097087 | 1 | 4.1875 |
| 28 | N | 0.219512 | 0.04878 | 4.5 | 3.83 |
| 29 | N | 0.220588 | 0.147059 | 1.5 | 3.75 |
| 30 | N | 0.305344 | 0.050891 | 6 | 3.22 |
| 31 | N | 0.306122 | 0.204082 | 1.5 | 2.8125 |
| 32 | N | 0.401606 | 0.040161 | 10 | 2.56 |
| 33 | N | 0.401198 | 0.299401 | 1.34 | 2.1875 |
| 34 | N | 0.5 | 0.04 | 12.5 | 2.11 |
| 35 | N | 0.5 | 0.238095 | 2.1 | 1.8125 |
| 36 | C | 0.097561 | 0.04878 | 2 | 3.95 |
| 37 | C | 0.097087 | 0.097087 | 1 | 3.823529 |
| 38 | C | 0.219512 | 0.04878 | 4.5 | 3.45 |
| 39 | C | 0.220588 | 0.147059 | 1.5 | 3.588235 |
| 40 | C | 0.305344 | 0.050891 | 6 | 2.85 |
| 41 | C | 0.306122 | 0.204082 | 1.5 | 2.529412 |
| 42 | C | 0.401606 | 0.040161 | 10 | 2.15 |
| 43 | C | 0.401198 | 0.299401 | 1.34 | 2.117647 |
| 44 | C | 0.5 | 0.04 | 12.5 | 1.75 |
| 45 | C | 0.5 | 0.238095 | 2.1 | 1.823529 |

TABLE IV
PEARSON CORRELATION COEFFICIENT (*r*)

| | | Pause Frequency | Pause Duration | Pause Intensity |
|---|---|---|---|---|
| Subjective Testing -1 | Moto GP (M) | -0.040 | -0.760 | -0.953 |
| | Run (R1) | -0.316 | -0.505 | -0.972 |
| | News (N) | -0.470 | -0.381 | -0.973 |
| | Cartoon (C) | -0.355 | -0.499 | -0.979 |
| Subjective Testing -2 | Rally (R2) | -0.366 | -0.254 | -0.923 |

In addition, the Pearson Correlation Coefficient or simply correlation coefficient (*r*) is also adopted to evaluate the correlation performance of PI, pause frequency and pause duration with MOS, respectively, as shown in Table IV. The correlation coefficient, *r*, for all the types of video sequences is calculated based on the results in Table III and Table V, and using the formula given below [28].

$$r = \frac{\sum(x - \bar{x})(y - \bar{y})}{\sqrt{\sum(x - \bar{x})^2 \sum(y - \bar{y})^2}} \quad (17)$$

where *x* and *y* are two sets of data, one representing PI, pause frequency, or pause duration and the other representing MOS; $\bar{x}$ and $\bar{y}$ are the mean of *x* and *y*, respectively. The correlation coefficient is used here to examine the linear relationship between PI, pause frequency or pause duration and MOS.

The correlation coefficients in Table IV clearly show that PI has a consistent high correlation with the viewer's experienced quality while the performances of pause frequency and pause duration are inconsistent and they have low correlation in most cases, especially for pause frequency. We have also examined the Spearrman's rank correlation coefficient for this work and found that it gives very similar results to those produced by applying the Pearson correlation coefficient.

### C. Subjective Assessment-2 for Pause Intensity

The second subjective testing session allows for stress testing of the pause intensity metric, where similar PI values are used with vastly different compositions. As shown in Table V, the bolded values represent the sequences with a high pause frequency and shorter pause duration while the opposite scenarios are represented by the non bolded values. Given these extreme cases of pause frequency and pause duration combinations whilst maintaining similar PI values, the aim of this testing session is to explore the suitability of pause intensity in these scenarios. Based on Table V and the resulting Fig.13, we are able to assess the correlation performance of various buffer characteristics with the perceived viewer quality (MOS), in the same way as for testing session 1.

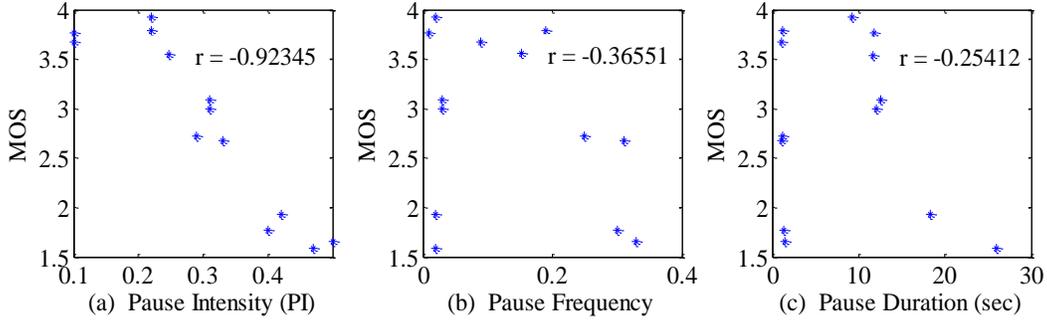

Fig. 13. Results for Subjective Testing-2: MOS vs. (a) PI, (b) pause frequency, and (c) pause duration.

Again, it is evident from Fig. 13(a) and Table V that a clear correlation between MOS and PI can be obtained even with the huge difference in pause characteristics composition. For example, both Videos 2 and 3 in Table V have the same PI value of 0.22 and receive very similar MOS ratings, in spite of the very different pause frequency and duration compositions in these cases. The respective average pause frequency in Video 3 is 0.19, much higher than that in Video 2 which is just 0.02; while the average pause duration of Video 3 is around 1.2 seconds, much shorter than 9.3 seconds for Video 2. This property is also agreed by the Pearson Correlation Coefficient indicated.

Fig. 13(b) shows, however, that the quality of experience of viewers as per MOS is inconsistent with the values of pause frequency. Videos 2 and 10, for example, have the same pause frequency (0.02), but their MOS values (3.93 and 1.50) do not match by a big margin, according to Table V.

Fig. 13(c) also indicates the shortcomings of using pause duration as a quality metric. Five videos (1, 3, 4, 7, 8) all have pause durations of around 1.1 seconds but the viewers' quality of experience varies greatly. It is also noticed that although the MOS ratings of Videos 0 and 1 are very close (3.76 and 3.67), their average pause durations are so different (11.76 and 1.08), showing no correlation between them.

Looking back at Table IV, the correlation coefficient for PI in Subjective Testing-2 overwhelmingly outperforms the other two pause characteristics. In addition, although pause duration has shown some correlation with varied levels in Subjective Testing-1, this performance is inconsistent with testing scenarios such as in Subjective Testing-2 where a very low correlation is recorded for pause duration.

All the results have demonstrated that PI performs consistently and holds a linear relationship with MOS in varied testing scenarios and this relationship is content independent.

TABLE V
SUBJECTIVE TESTING-2 RESULTS

| Video ID | PI | Pause Frequency | Average Pause Duration | MOS |
|---|---|---|---|---|
| 0 | 0.10 | 0.01 | 11.76 | 3.76 |
| 1 | 0.10 | 0.09 | **1.08** | **3.67** |
| 2 | 0.22 | 0.02 | 9.29 | 3.93 |
| 3 | 0.22 | 0.19 | **1.17** | **3.79** |
| 4 | 0.29 | 0.25 | **1.17** | **2.72** |
| 5 | 0.31 | 0.03 | 12.00 | 3.00 |
| 6 | 0.31 | 0.03 | 12.52 | 3.09 |
| 7 | 0.33 | 0.31 | **1.08** | **2.68** |
| 8 | 0.40 | 0.30 | **1.33** | **1.77** |
| 9 | 0.42 | 0.02 | 18.32 | 1.93 |
| 10 | 0.47 | 0.02 | 25.98 | 1.59 |
| 11 | 0.50 | 0.33 | **1.50** | **1.65** |

## VII. CONCLUSION

This paper has explored the pause intensity metric in the context of video streaming in a TCP network, and provided a detailed analysis of the advantages of PI over other existing metrics that characterize buffer underrun. The analytical model developed reveals that the PI metric can be determined by both the video playout rate and network throughput and, more precisely, it is a ratio of the rate difference ($\lambda$-$\eta$) to the playout rate $\lambda$.

We have verified the model through extensive simulation using NS-2, which demonstrates the high accuracy of the model established. We have also provided subjective testing results based on 57 video streaming clips, signifying that the PI metric has a very good correlation with the viewer's quality of experience (QoE) in terms of the MOS ratings. In addition, we have shown that either pause frequency or pause duration is not sufficient on its own to reflect the perceived video playback quality by viewers. The results from subjective tests have confirmed the independency and consistency of our metric through testing different types of video clips and allowing vastly different compositions (constituted by both pause frequency and pause duration) for defining PI values.

Being a no-reference metric and related to network performance and system settings, PI values can be easily made available for network operators and services providers to predict user's QoE even before the video is actually played out at the receiver. As such, the conventional control mechanism such as rate adaptation can then be used more effectively and in a quality-guided manner. Further exploitation of this work will involve applying the higher-order factor analysis to the PI metric and realizing its benefit for real systems.